\documentstyle[pra,preprint,psfig,aps]{revtex}
\psfigurepath{/month/saif/Part2/latex/}

\begin{document}
\draft
\title{Quantum Recurrences: Probe to study Quantum Chaos}
\author{Farhan Saif}
\address{Department of Electronics, Quaid-i-Azam university, 54320\\
Islamabad, Pakistan.}
\maketitle

\begin{abstract}
We study the phase space of periodically modulated gravitational cavity by
means of quantum recurrence phenomena. We report that the quantum
recurrences serve as a tool to connect phase space of the driven system with
spectrum in quantum domain. With the help of quantum recurrences we
investigate the quasi-energy spectrum of the system for a certain fixed
modulation strength. In addition, we study transition of spectrum from
discrete to continuum as a function of modulation strength.
\end{abstract}


%\begin{multicols}{2}

What are the generic properties of quantum chaos which provide a firm
understanding of chaos in quantum mechanical domain? In the young field of
quantum chaos this question has occupied the researchers right from the very
beginning~\cite{kn:haak,kn:hogg,kn:bres,kn:toms,kn:auri}. In this paper we
suggest a partial answer to this question. We study the wavepacket dynamics
in a periodically driven system and probe the chaotic phase space by means
of quantum recurrences.

Discreteness of quantum mechanics manifests itself in the phenomenon of
quantum recurrences~\cite{kn:Eberly}. In one degree of freedom systems
quantum recurrences have been studied ~\cite{kn:aver} and has been applied
in vast variety of subjects from femto-second chemistry~\cite{kn:Zewail} to
isotope separation~\cite{kn:aver1,kn:saifi}. In higher degree of freedom
systems study of quantum recurrences is a new subject~\cite{kn:saif1}. The
presence of quantum recurrences in some such systems has been pointed out
earlier\ \ \cite{kn:haak,kn:hogg,kn:bres,kn:toms,kn:auri,kn:saif1,kn:saif2}.
We establish numerically that the phenomena of quantum recurrences or
quantum revivals together with fractional revivals are generic to the higher
dimensional systems exhibiting quantum chaos. Moreover, by probing classical
phase space with the help of quantum revival phenomena we report that ({\it i%
}) quantum evolution is different for different initial conditions, and (%
{\it ii}) quantum revivals carry the information of underlying quasi-energy
spectrum.

In this paper, we consider the dynamics of cold atoms moving under the
influence of gravity and bouncing off an evanescent wave mirror~\cite
{kn:wallis}. We provide an external periodic modulation to the mirror by
means of an acusto-optic modulator~\cite{kn:sten}. The Schr\"{o}dinger
equation, 
\begin{equation}
ik^{\hspace{-2.1mm}-}\frac{\partial \psi }{\partial t}=\left[ \frac{p^{2}}{2}%
+z+V_{0}\exp \left[ -\kappa (z-\lambda \sin t)\right] \right] \psi ,
\label{eq:tdsam}
\end{equation}
\noindent controls the dynamics of an atom moving in the modulated
gravitational cavity~\cite{kn:saif3}. The dimensionless coordinates $z$ and $%
p$ are scaled by using the frequency of the external modulation $\omega $,
mass of the atom $M$, and gravitational constant $g$ as $z=\tilde{z}\omega
^{2}/g$, $p=\tilde{p}\omega /Mg$, where $\tilde{z}$ and $\tilde{p}$ are real
coordinates. These scaled coordinates satisfy the commutation relation $%
[z,p]=\omega ^{3}/Mg^{2}[\tilde{z},\tilde{p}]=i\hbar \omega ^{3}/Mg^{2}=ik^{%
\hspace{-2.1mm}-}$. Here, $V_{0}$ and $\kappa $ indicate the height and the
steepness of the exponential potential, respectively. Moreover, we express
the dimensionless strength of the external spatial modulation, provided by
an acusto-optic modulator to the gravitational cavity, by $%
\lambda=a\omega^2/g$, where $a$ is the amplitude of the modulation.

The classical evolution of this system follows Liouville equation~\cite
{kn:saif1}. We study the classical dynamics with the help of Poincar\'e
surface of section obtained for 25 atoms propagated in gravitational cavity
in the presence of an external modulating field of strength $\lambda=0.3$,
as shown in Fig.~\ref{fg:pserev}. In the upper right corner of the
Poincar\'e section, we show area of a unit cell in quantum domain.

For the sake of clarity, we have investigated the quantum evolution for two
different sets of initial conditions in phase space: First set comprises $%
\{(14.5, 1.45),(15,0),(15,-1),(15,-2)\}$. We label these phase points as $a,
b, c, d$, respectively. In phase space we represent these points as centers
of circles approximately around $z=15$ line, as in Fig.~\ref{fg:pserev}.
Phase space point $a$ sits, approximately, at the center of primary
resonance 2:1, and $b$ at the edge of the same resonance. The other two
phase points $c$ and $d$ correspond to the stochastic sea. We have confirmed
the location of phase points by calculating the Lyapunov exponent. For $a$
and $b$ we find zero Lyapunov exponent, whereas in case of $c$ and $d$, we
have non-zero positive exponents which show an exponential divergence for
these initial conditions. The second set of phase space points investigate
the effect of secondary resonances on quantum dynamics in the modulated
gravitational cavity. For this purpose we choose our initial conditions as $%
\{(10,0), (25,0)\}$, which are at the left and at the right of $z=15$ line,
respectively, and again express centers of circles, as shown in Fig.~\ref
{fg:pserev}. We label them as $e$ and $f$. The Lyapunov exponents,
corresponding to these phase points, are zero.

We propagate an initially well localized atomic wavepacket, $\psi(0)$,
starting from each of the phase point of the two sets and note its
evolution. The initial size of the atomic wavepacket satisfies minimum
uncertainty relationship and is expressed by the circles around each of the
phase point, as shown in Fig. 1. In order to study the dynamics of the
wavepacket in the modulated gravitational cavity we calculate square of
auto-correlation function, 
\begin{equation}
C^2=|\langle\psi(0)|\psi(t)\rangle|^2,
\end{equation}
where $\psi(t)$ is the atomic wavefunction after an evolution time, $t$, in
the driven system.

In the absence of any modulation, that is for $\lambda=0$, the wavepacket
displays well investigated quantum revivals for one degree of freedom
systems~\cite{kn:aver,kn:chen} for all initial conditions. As we switch on
the external modulation, the net system comprises two degrees of freedom. We
find that the behavior of revival phenomena changes drastically. We find a
complete disappearance of the quantum revivals for the atomic wavepacket
originating approximately around the center of the primary resonance, $a$.
In contrast, this atomic wavepacket displays almost a complete revival after
classical revival time, as shown in Fig.~\ref{fg:rev12}(a). We calculate the
classical revival time as $T_{cl}=4\pi$ by approximating the potential of
the gravitational cavity by triangular well potential~\cite{kn:saif1}. The
analytical result agrees well with the numerically obtained classical period.

We may understand this interesting property by noting that a resonance can
be expressed effectivly by pendulum Hamiltonian~\cite{kn:saif1}, 
\begin{equation}
H= -\frac{\partial^2}{\partial\varphi^2}+ V_0\cos\varphi.
\end{equation}
Therefore, when $\varphi\ll 1$, the effective Hamiltonian of the system is 
\begin{equation}
H\approx -\frac{\partial^2}{\partial\varphi^2}- \frac{V_0}{2}\varphi^2,
\end{equation}
of a harmonic oscillator. This effective Hamiltonian controls the evolution
of an atomic wavepacket placed close to the center of the resonance. This
analogy provides us an evidence that if an atomic wavepacket is placed
initially around the center of a resonance it will always observe revivals
after each classical period, as in case of a harmonic oscillator.

In addition, this analogy provides an information about level spacing around
the center of a resonance in the driven gravitational cavity. Since in case
of harmonic oscillator the spacing between successive levels is always
equal, we conclude that the spacing between quasi-energy levels is equal
around the center of resonance in a periodically driven system.

If we place the atomic wavepacket at an edge of the primary resonance, it
follows classical trajectory in its early evolution, and exhibits periodic
recurrences after the classical period, $T_{cl}$. However, in the long time
dynamics we observe the emergence of the quantum revivals, as we show in
Fig.~\ref{fg:rev12}(b). We explain this behavior in the light of our earlier
discussion, that is, away from the center of the primary resonance, the
nonlinearity of the original potential contributes to the effective harmonic
potential of the resonance. As a result, we find the appearance of quantum
revivals in presence of external modulation, together with the classical
periodic motion.

In order to elaborate the effect of stochastic region we propagate the
atomic wavepacket centered at the phase points $c$ and $d$. On calculating
the square of the autocorrelation function for the two initial conditions,
we find that even after a time much larger than Ehrenfest's time, there
occurs no revival phenomena, as shown in Fig.~\ref{fg:rev12}(c) and 2(d). We
conjecture that in the stochastic region the quasi-energy spectrum makes a
quasi-continuum, therefore, causing the absence of quantum recurrences.

In order to provide a detailed investigation of revival phenomena we take
the phase points $e$ and $f$ which belong to the secondary resonances at the
left and at the right of z=15 line, respectively, as shown in Fig.~\ref
{fg:pserev}. We keep all the parameters the same as before and propagate the
wavepackets centered at these phase points. At the phase space point $e$,
the initial wavepacket sits mostly inside the island region. The quantum
revivals occur but the process of collapse and then revival of the atomic
wavepacket is rather slow. However, for the phase point $f$ the size of the
initial wavepacket is of the order of the stable island and therefore the
effect of the nonlinearity is more significant than the earlier case of
phase space point, $e$. Hence, we see that the wavepacket initiating from
this initial condition has very pronounced collapses and revivals, as we
display in Fig.~\ref{fg:rev34}.

We~\cite{kn:saif1,kn:saif2} may calculate the time of the quantum revivals
as 
\begin{equation}
T_{\lambda}=T_0\left[ 1-\frac{1}{8}\left\{ \frac{\lambda}{E_0} \right\}^{2} 
\frac{3(1-r)^2+a^2}{((1-r)^2-a^2)^3}\right],  \label{eq:8}
\end{equation}
where $r\equiv(E_N/E_0)^{1/2}$ and $a\equiv r^2 k^{\hspace{-2.1mm}-}/4E_0$.
Here, $T_0$ corresponds to the time of revival \cite{kn:chen} 
\begin{equation}
T_0= \frac{16 E^2_0}{\pi k^{\hspace{-2.1mm}-}},  \label{eq:lzrt}
\end{equation}
in the absence of any modulation. Moreover, $E_0$ is the average energy of
the initially excited wavepacket and $E_N$ is the energy of the $N$th
resonance.

Looking at the different behaviors of quantum revivals for the wavepackets
originating from different initial conditions, we conclude that the quasi
energy spectrum possesses a quasi-continuum structure in the stochastic
region of phase space causing the disappearance of revival structures.
However, we find a local discrete spectrum in the region of resonances
leading to periodic revivals and collapses of the wavefunction. Therefore,
the discrete spectrum at zero modulation develops band structures in
presence of external modulating field comprising local quasi continuum
separated by discrete levels.

Now, we come across another interesting question: What happens to the
quantum revivals of a driven system by varying the strength of the external
modulation? As we discussed earlier, in the absence of any external
modulation we find revival phenomenon for all the initial conditions. We can
calculate the corresponding revival time for undriven gravitational cavity
from Eq.~(\ref{eq:lzrt}). As we switch on the modulation these revivals
change significantly depending upon the initial condition of the propagated
wavepacket in phase space. From our numerical investigations, we find that
the atomic wavepacket placed around the center of a resonance shows almost
complete revival after each classical period. Thus the initial phenomenon of
quantum revivals which occurs for $\lambda=0$, disappears completely in
presence of non-zero modulation and the wavepacket displays almost a
complete recurrence after a classical period. In case the initial wavepacket
is around a separatrix, the revival phenomena occur only for very small
modulation strength, $\lambda\approx 0$, and vanish abruptly going beyond
these values and we do not see any recurrences at all.

In order to study general modification of the revival phenomena as a
function of modulation strength $\lambda$, we calculate the square of the
auto correlation function for the wavefunction originating from the phase
point $f$ and study its change with increasing modulation strength. We find
that in presence of the external modulation the revival time reduces with
the rising modulation, as shown in Fig.~\ref{fg:revlam}. We can calculate
revival time for the driven system using Eq.~(\ref{eq:8}). We find that the
revival structures survive together with the fractional revivals for smaller
values of the external modulation. However, on increasing the modulation
strength, $\lambda$, first the fractional revivals and then the quantum
revivals reduce in magnitude.

In the modulated gravitational cavity above a critical value of the
modulation strength $\lambda=\lambda_u$, quantum diffusion sets in~\cite
{kn:saif3}. At this critical value the spectrum of the system undergoes a
phase transition and changes from point spectrum to a continuum spectrum~ 
\cite{kn:bren,kn:benv,kn:oliv,kn:chen1}, and as a consequence, we find
quantum diffusion. We can identify this transition of the spectrum by noting
that the quantum revivals disappear completely as the modulation strength
exceeds the critical modulation strength.

By probing phase space with the help of revival phenomena, we conjecture
that the quantum mechanical initially discrete spectrum of the un-modulated
system changes to a band structure in presence of external modulation. It
keeps the discreteness in the vicinity of resonance with almost equal level
spacing at the center, and develops a quasi-continuum in stochastic region.
However, level spacing gradually reduces with the rising modulation and
disappears completely above the quantum diffusion limit, {\it i.e.} $%
\lambda=\lambda_u$. Hence, we find a change in the spectrum from discrete
spectrum at $\lambda =0$, to band spectrum for modulation strength smaller
than the critical modulation strength $\lambda _{u}$, and then to continuum
spectrum above $\lambda _{u}$. In this way we can probe all the three
regimes of the spectrum by looking at the revival phenomena of the atomic
wavepacket as a function of modulation strength. Moreover, the revival
structures also help to differentiate the local quasi continuum from local
discrete spectrum occurring for modulations smaller than the critical
modulation strength.

We thank G. Alber, M.\ Fortunato, R.\ Grimm, B. Mirbach, W. P. Schleich, F.
Steiner, and M. S. Zubairy for many fruitful discussions.

%\end{multicols}

\vspace{1ex} 
\begin{figure}[tbp]
\caption{The Poincar\'{e} surface of section for the modulation amplitude $%
\lambda =0.3$: The phase space displays overlap of the resonances. The
centers of the circles correspond to the chosen phase points and the size of
each circle corresponds to the size of the wavepacket. In the right upper
corner we display the unit size, which is $2\pi k^{\hspace{-2.1mm}-}$, of
the quantum space by the dark box. }
\label{fg:pserev}
\end{figure}
\vspace{1ex}

\vspace{1ex} 
\begin{figure}[tbp]
\caption{The change in revival phenomena for the wavepacket originating from
different initial condition in the phase space: We display the revival
structures of the initial Gaussian wavepacket for a modulation strength $%
\lambda=0.3$. The initial wavepacket originates from the (a) center of the
primary resonance (14.5, 1.45) and from the phase points (b) (15,0), (c)
(15,-1), (d) (15,-2). For the wavepacket originating at the center of the
resonance we find revivals after classical period, whereas, for the
wavepacket sitting initially at the edge of the resonance we find the
quantum revivals. However, the revival structures disappear if the
wavepacket originates from the stochastic region, as we find in case (c) and
(d). We have considered $V_0=1$, $\kappa=1$ and the effective Planck's
constant as $k^{\hspace{-2.1mm}-}=1$. }
\label{fg:rev12}
\end{figure}
\vspace{1ex}

\vspace{1ex} 
\begin{figure}[tbp]
\caption{Comparison of revivals for the wavepacket originating from
secondary resonances: (top) The wavepacket originates from the phase points
(25,0) and (bottom) (10,0). We kept all the parameters the same as in Fig.~%
\ref{fg:rev12}.}
\label{fg:rev34}
\end{figure}
\vspace{1ex}


\begin{references}
\bibitem{kn:haak}  F. Haake, {\it Quantum Signatures of Chaos}, (Springer,
Berlin 1992).

\bibitem{kn:hogg}  T. Hogg and B. A. Huberman, Phys. Rev. Lett. {\bf 48},
711 (1982).

\bibitem{kn:bres}  J. K. Breslin, C.A. Holmes, and G.J. Milburn, Phys. Rev.
A {\bf 56}, 3022 (1997).

\bibitem{kn:toms}  S. Tomsovic and J. Lefebvre, Phys. Rev. Lett. {\bf 79},
3629 (1997).

\bibitem{kn:auri}  R. Aurich and F. Steiner, Int. J. Mod. Phys. B {\bf 13},
2361 (1999).

\bibitem{kn:Eberly}  J. H. Eberly, N. B. Narozhny, and J. J.
SanchezMondragon, Phys. Rev. Lett. {\bf 44}, 1323 (1980).

\bibitem{kn:aver}  I. Sh. Averbukh and N.F. Perel'man, Phys. Lett. A {\bf 139%
}, 449 (1989); P. A. Braun and V. I. Savichev, J. Phys. B {\bf 29}, L329
(1996); C. Leichtle, I. Sh. Averbukh and W. P. Schleich, Phys Rev. Lett {\bf %
77}, 3999 (1996); {\it ibid}, Phys Rev. A {\bf 54}, 5299 (1996).

\bibitem{kn:Zewail}  A. H. Zewail, {\it Femtochemistry} (World Scientific,
Singapore, 1994), Vols. 1 and 2.

\bibitem{kn:aver1}  I. Sh. Averbukh, Marc J. J. Vrakking, D. M. Villeneuve
and A. Stolow, Phys. Rev. Lett. {\bf 77}, 3518 (1996).

\bibitem{kn:saifi}  F. Saif, to be published.

\bibitem{kn:saif1}  F. Saif, Ph.D thesis (Ulm Universit\"{a}t, Ulm, 1998).

\bibitem{kn:saif2}  F. Saif, G. Alber, V. Savichev, and W.P. Schleich, to be
submitted.

\bibitem{kn:wallis}  H. Wallis, J. Dalibard, and C. Cohen-Tannoudji, Appl.
Phys. B {\bf 54}, 407 (1992).

\bibitem{kn:sten}  A. Steane, P. Szriftgiser, P. Desbiolles and J. Dalibard,
Phys. Rev. Lett. {\bf 74}, 4972 (1995).

\bibitem{kn:saif3}  F. Saif, I. Bialynicki-Birula, M. Fortunato, and W.P.
Schleich, Phys. Rev. A 4779, {\bf 58} 1998.

\bibitem{kn:chen}  Wen-Yu Chen and G.J. Milburn, Phys. Rev. A {\bf 51}, 2328
(1995).

\bibitem{kn:blum}  R. Bl\"{u}mel and W. P. Reinhardt, {\it Chaos in atomic
Physics}, (Cambridge, New York, 1997).

\bibitem{kn:bren}  N. Brenner and S. Fishman, Phys. Rev. Lett. {\bf 77},
3763 (1996).

\bibitem{kn:benv}  F. Benvenuto, G. Casati, I. Guarneri and D.L.
Shepelyansky, Z. Phys. B {\bf 84}, 159 (1991);

\bibitem{kn:oliv}  C.R. de Oliveira, I. Guarneri and G. Casati, Europhys.
Lett. {\bf 27}, 187 (1994).

\bibitem{kn:chen1}  Wen-Yu Chen and G. J. Milburn, Phys. Rev. E {\bf 56},
351 (1997).
\end{references}
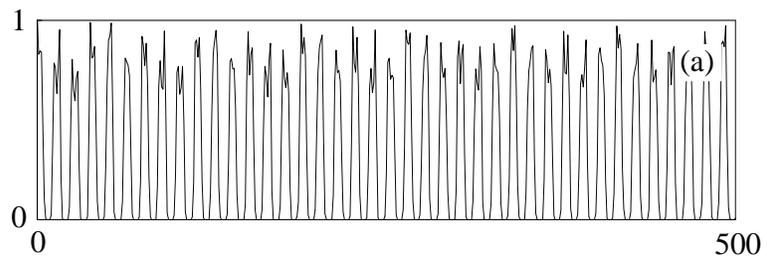
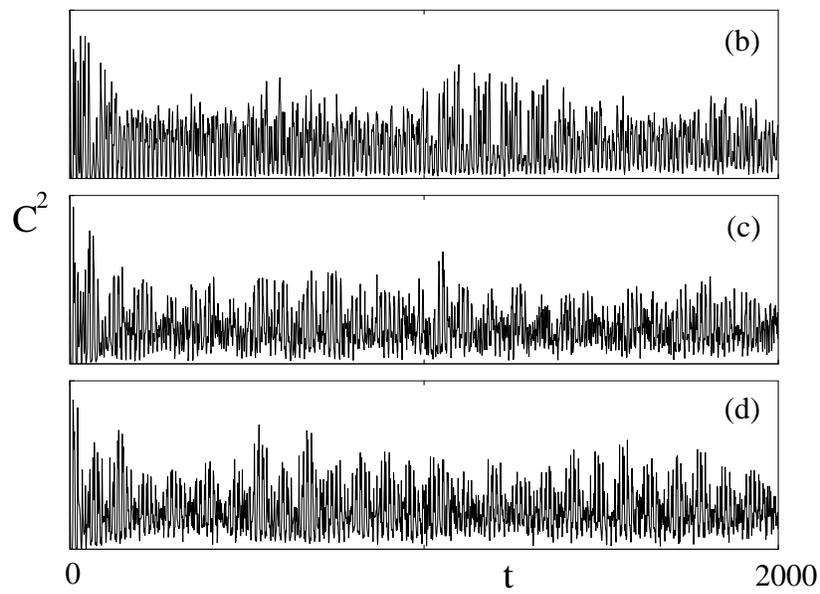
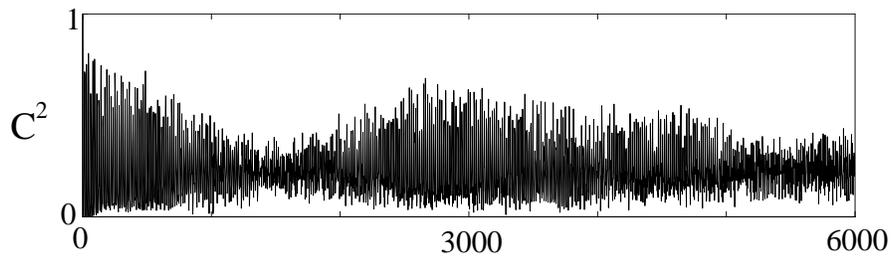
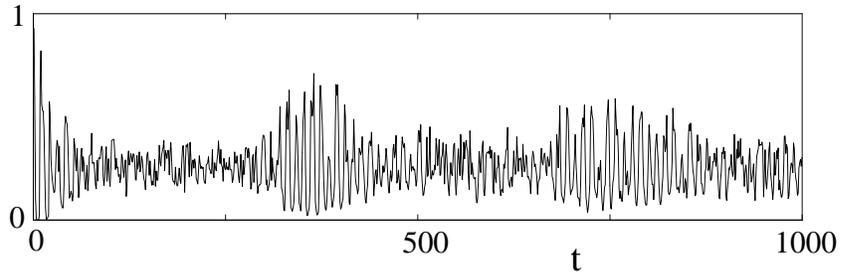
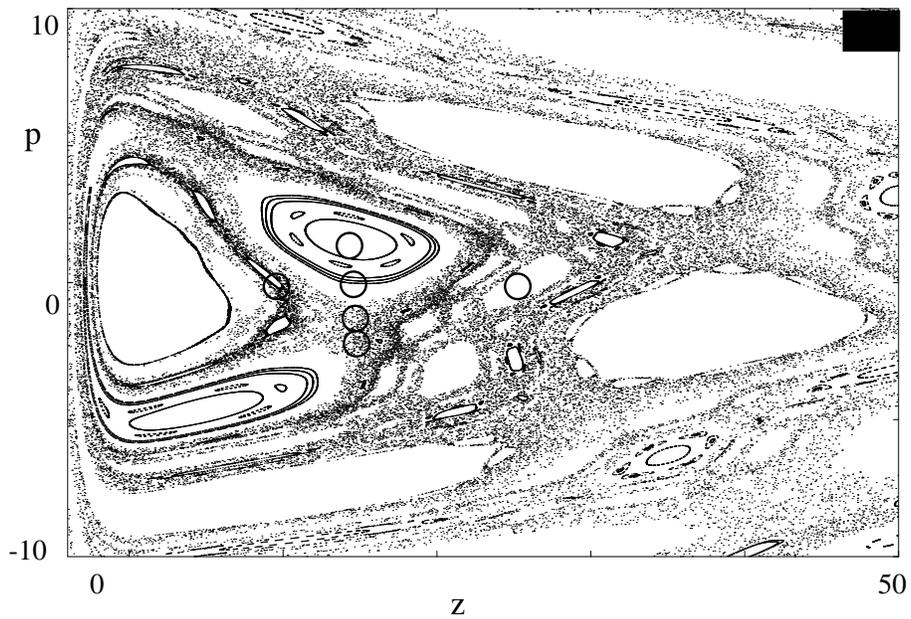
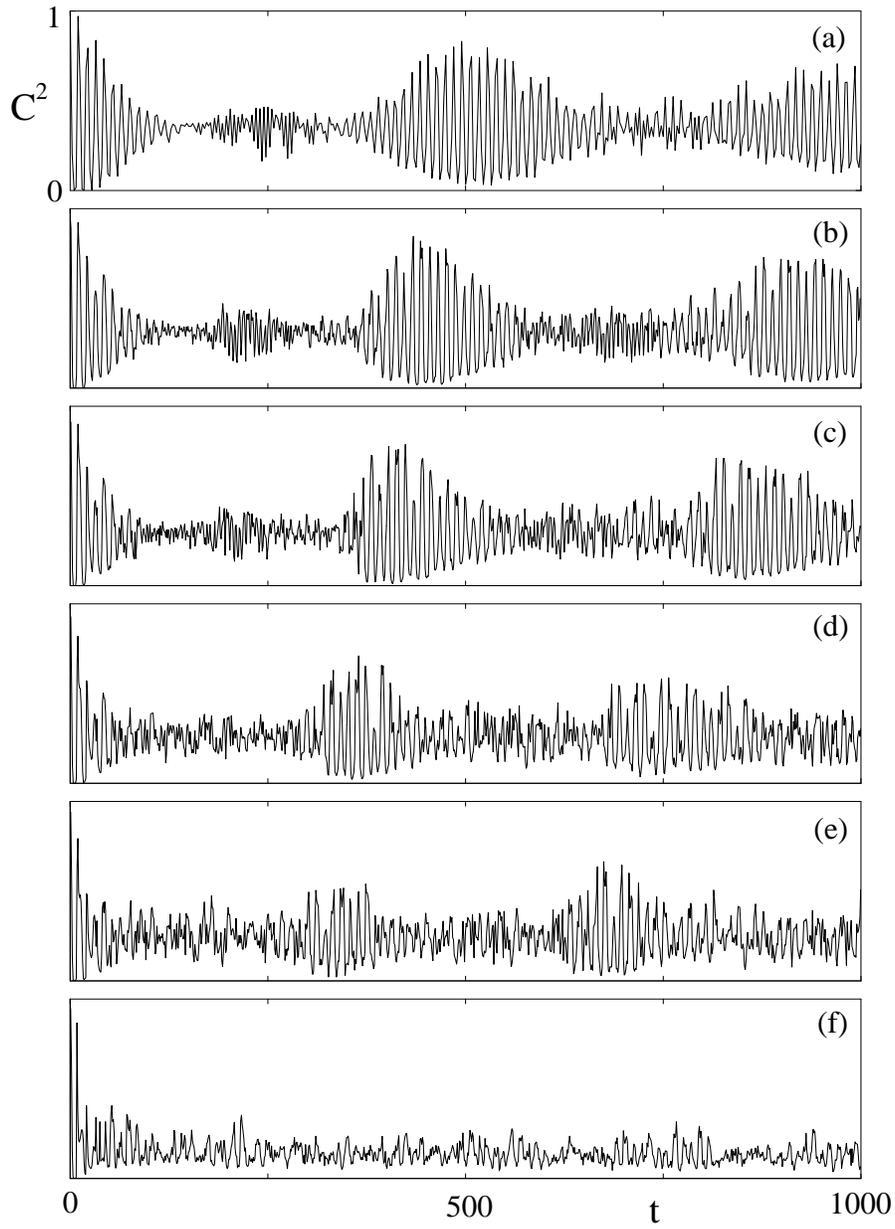
\end{document}